\documentclass[aps,prl,twocolumn,superscriptaddress]{revtex4-1}
\usepackage{graphicx}
\usepackage[german,english]{babel}
\usepackage{amssymb}
\usepackage{amsmath}
\usepackage{braket}
\usepackage{subfigure}

\begin{document}

\title{ Orbital Kondo effect in Cobalt-Benzene sandwich molecules }

\author{M. Karolak}
\affiliation{I. Institut f{\"u}r Theoretische Physik, Universit{\"a}t Hamburg, 
Jungiusstra{\ss}e 9, D-20355 Hamburg, Germany}

\author{D. Jacob}
\affiliation{Max-Planck-Institut f{\"u}r Mikrostrukturphysik, Weinberg 2, 
06120 Halle, Germany}

\author{A. I. Lichtenstein}
\affiliation{I. Institut f{\"u}r Theoretische Physik, Universit{\"a}t Hamburg, 
Jungiusstra{\ss}e 9, D-20355 Hamburg, Germany}


\date{\today}

\begin{abstract}
We study a Co-benzene sandwich molecule bridging the tips of a Cu nanocontact 
as a realistic model of correlated molecular transport. 
To this end we employ a recently developed method for calculating the correlated 
electronic structure and transport properties of nanoscopic conductors.
When the molecule is slightly compressed by the tips of the nanocontact the dynamic 
correlations originating from the strongly interacting Co $3d$ shell give rise to an 
orbital Kondo effect while the usual spin Kondo effect is suppressed due to Hund's 
rule coupling.
This non-trivial Kondo effect produces a sharp and temperature-dependent Abrikosov-Suhl 
resonance in the spectral function at the Fermi level and a corresponding Fano line shape 
in the low bias conductance.
\end{abstract}

\maketitle

The discovery of ferrocene and bisbenzene chromium \cite{kealy1951,*fischer1955}
over half a century ago was the starting point for experimental and theoretical work 
both in chemistry and physics concerning the intricate properties of organometallic 
compounds. The investigation of these and related sandwich complexes is driven by their 
relevance in various chemical applications (e.g. catalysis) and more recently also 
because of their prospective nanotechnological applications for example as molecular 
magnets \cite{gatteschi} or spintronic devices \cite{xiang2006,*mokrousov2007}. 

Molecules with a transition metal (TM) center coupled to aromatic groups are also of 
high interest from a fundamental point of view. The strong electronic correlations in 
the $d$ shell of the TM can modify the ground state and electronic 
transport properties of such molecules, leading to many-body phenomena like the Kondo 
effect \cite{hewson} as recently observed in TM-phthalocyanine molecules 
\cite{zhao2005,*gao2007}.
The importance of dynamical correlations in nanoscopic devices in general is further 
substantiated by the recent observation of the Kondo effect in nanocontacts made from 
TMs \cite{calvo2009,jacob2009}. Also recently a so-called underscreened Kondo effect
has been reported for a molecule trapped in a breakjunction \cite{parks2010}.
While the Kondo effect is commonly associated with the screening of a local magnetic 
moment by the conduction electrons, it is also possible that another internal degree 
of freedom associated with a degeneracy gives rise to a Kondo effect \cite{cox1998}. 
One example is the so-called {\it orbital Kondo effect} 
where the pseudo-spin arising from an orbital degeneracy is screened
by the conduction electrons \cite{kolesnychenko2002,*jarillo2005}.

Such complex many-body phenomena call for a theoretical description beyond the standard 
treatment with Kohn-Sham (KS) density functional theory (DFT) which cannot capture 
many-body physics beyond the effective mean-field picture. 
The incorrect behavior of the KS DFT for strongly correlated systems can be 
remedied by augmenting the DFT with a local Hubbard-like interaction.
This DFT++ approach is the de facto standard in the 
theory of solids \cite{kotliar_rmp,*lda++}. Recently this approach has been 
adapted to the case of nanoscopic conductors 
\cite{jacob2009,jacob2010_1,*jacob2010_2,lucignano2009,*dasilva2009,*korytar2011,*valli2010}.

In this letter we apply this method to investigate the transport 
properties of cobalt benzene sandwich molecules (C$_6$H$_6$)Co(C$_6$H$_6$) 
(CoBz$_2$ in the following). We find that dynamical correlations in the Co 
$3d$ shell give rise to an {\it orbital Kondo effect} in a doubly degenerate 
level of the Co $3d$ shell while the usual spin Kondo effect is suppressed 
due to Hund's rule coupling.

We perform DFT calculations using the CRYSTAL06 code \cite{crystal} employing the LDA 
\cite{kohn_sham}, PW91 \cite{pw91} and the hybrid functional B3LYP \cite{becke}, 
and using the all electron Gaussian 6-31G basis set.
The geometries of the wires were relaxed beforehand and kept fixed during the calculations. 
The geometry of the molecule in contact with the wires was relaxed employing the B3LYP 
functional \cite{geometry}.

In order to capture many-body effects beyond the DFT level, we have applied the 
DFT+OCA (Density Functional Theory + One-Crossing Approximation) for nanoscopic 
conductors developed by one of us in earlier work \cite{jacob2009,jacob2010_1}. 
To this end the system is divided into three parts: The two semi-infinite leads 
L and R and the device region D containing the molecule and part of the leads.
The KS Green's function (GF) of region D can now be obtained from the DFT 
electronic structure as
$G^0_{\rm D}(\omega)=(\omega+\mu-H^0_{\rm D}-\Sigma_{\rm L}(\omega)-\Sigma_{\rm R}(\omega))^{-1}$
where $H^0_{\rm D}$ is the KS Hamiltonian of region D and $\Sigma_{\rm L,R}(\omega)$ are 
the so-called lead self-energies which describe the coupling of D to L and R 
and which are obtained from the DFT electronic structure of the nanowire leads. 

Next the mean-field KS Hamiltonian is augmented by a Hubbard-like interaction term 
$\sum_{ijkl} U_{ijkl}\hat{d}_i^\dagger\hat{d}_j^\dagger\hat{d}_l\hat{d}_k$ that accounts 
for the strongly interacting electrons of the Co $3d$ shell. Here we use a simplified 
interaction which only takes into account the direct Coulomb repulsion $U\equiv U_{ijij}$ 
and the Hund's rule coupling $J\equiv U_{ijji}$. These are different from 
the bare interactions due to screening processes. Here we assume that the interactions
are somewhat increased as compared to their bulk values since the screening should
be weaker than in bulk. Hence we use $U=5$~eV and $J=1$~eV; indeed we find that our results are 
qualitatively stable for a reasonable range of values for $U$ and $J$ (see below).

The interacting Co $3d$ shell coupled to the rest of the system (benzene+leads) 
constitutes a so-called Anderson impurity model (AIM). The AIM is completely defined by 
the interaction parameters $U$ and $J$, the energy levels $\epsilon_d$ of the $3d$ 
orbitals and the so-called hybridization function $\Delta_d(\omega)$.
The latter describes the (dynamic) coupling of the Co $3d$-shell to the rest
of the system and can be obtained from the KS GF as 
$\Delta_d(\omega)=\omega+\mu-\epsilon_d^0-[G^0_d(\omega)]^{-1}$
where $\mu$ is the chemical potential, $\epsilon_d^0$ are the KS energy levels 
of the $3d$ orbitals and $G^0_d(\omega)$ is the KS GF projected onto the $3d$ 
subspace. 
The energy levels $\epsilon_d$ are obtained from the static crystal field
$\epsilon_d^0$ where as usual in DFT++ approaches a double counting correction 
(DCC) has to be subtracted to compensate for the overcounting of interaction terms.
Here we employ the so-called fully localized or atomic limit DCC \cite{czyzyk1994} 
$E_{dc}=U\cdot\left(N_{3d}-\frac{1}{2}\right)-J\left(\frac{N_{3d}}{2}-\frac{1}{2}\right)$.

The AIM is solved within the OCA \cite{haule2001,kotliar_rmp}. 
This yields the electronic self-energy $\Sigma_d(\omega)$ which accounts for the electronic 
correlations of the $3d$-electrons due to strong electron-electron interactions. The 
{\it correlated} $3d$ GF is then given by $G_d=([G_d^0]^{-1}-\Sigma_d+E_{dc})^{-1}$. 
Correspondingly, the {\it correlated} GF for D is given by 
$G_{\rm D}=([G_{\rm D}^0]^{-1}-\Sigma_d+E_{dc})^{-1}$ where $\Sigma_d$ and $E_{dc}$
only act within the $3d$ subspace.
From $G_{\rm D}$ we can calculate the transmission function
$T(\omega)={\rm Tr}[\Gamma_{\rm L}G_{\rm D}^\dagger\Gamma_{\rm R}G_{\rm D}]$
where $\Gamma_{\alpha}\equiv i(\Sigma_{\alpha}-\Sigma_{\alpha}^\dagger)$. 
For small bias voltages $V$, the transmission yields the conductance:
$\mathcal{G}(V)=(2e^2/h)T(eV)$.

\begin{figure}
  \begin{center}
    \includegraphics[width=0.9\linewidth]{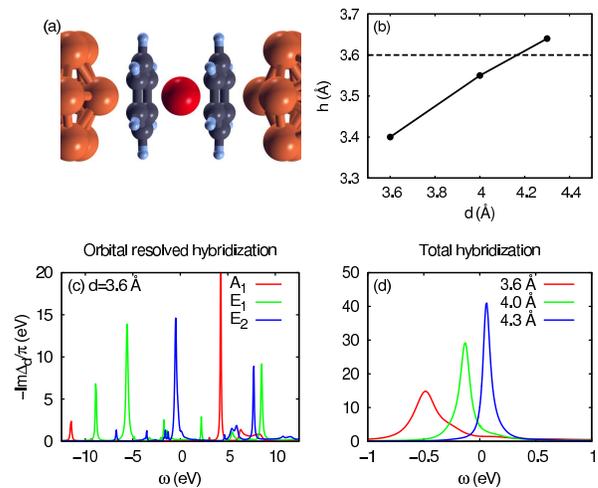}
  \end{center}
  \caption{(Color online)
    (a) Atomic structure of CoBz$_2$ molecule in a Cu nanocontact.
    (b) Size $h$ of CoBz$_2$ molecule as a function of the distance $d$
    between Co atom and Cu tip atoms.
    (c) Orbitally resolved imaginary part of the hybridization function for $d=3.6$\r{A}.
    (d) Total hybridization function (all Co $3d$ orbitals) for different $d$.
  }
  \label{dft-results}
\end{figure}

We consider a single  CoBz$_2$ molecule in contact with two semi-infinite Cu 
nanowires as shown in Fig.~\ref{dft-results}a. The CoBz$_2$ molecule is the smallest 
instance of a general class of M$_n$Bz$_{n+1}$ complexes, where M stands for a metal atom, 
that have been prepared and investigated 
\cite{kurikawa1999,*nakajima2000,*martinez2010,xiang2006,*mokrousov2007,*zhang_2008}. 
The semi-infinite Cu wires exhibit the hexagonal symmetry of the molecule and correspond 
to the (6,0) wires described in Ref. \onlinecite{tosatti2001}.
We investigate the system at three different Cu-tip-Co distances 3.6~\r{A}, 4.0~\r{A} and 4.3~\r{A} (see Fig.~\ref{dft-results}a,b).
As can be seen from Fig.~\ref{dft-results}b the Bz-Bz distance $h$ varies 
between 3.4~\r{A} and 3.65~\r{A} depending on the distance $d$ of the Cu tip to the Co 
atom in the center of the molecule. At distances $d=3.6$~\r{A} and $d=4.0$~\r{A} the 
molecule is slightly compressed compared to its \textit{free} height of about 
$h=3.6$~\r{A}, whereas it is slightly stretched at $d=4.3$~\r{A}. 

The hexagonal symmetry of the system leads to a lifting of the degeneracy 
of the five $3d$ orbitals. The symmetry adapted representations are: the $A_1$ group 
consisting of the $d_{3z^2-r^2}$ orbital only, the doubly degenerate $E_1$ group consisting 
of the $d_{xz}$ and $d_{yz}$ orbitals and the $E_2$ group consisting of the $d_{xy}$ and 
$d_{x^2-y^2}$ orbitals.

Figs. \ref{dft-results}c+d show the imaginary parts of the hybridization functions 
$\Delta_d(\omega)$ calculated from the LDA electronic structure. 
The imaginary part of the hybridization function exhibits a distinct peak close to the 
Fermi level ($E_F$) in the $E_2$ channel, whose position, width and height depend 
significantly on the molecular geometry, specifically on the Bz-Co distance. The other 
channels $E_1$ and $A_1$ show only a negligible hybridization close to $E_F$. 
The dominant feature stems, similarly as shown for graphene \cite{wehling2010,jacob2010_1} 
from hybridization with the $\pi_z$ orbital state of the benzene rings. The feature does 
not depend qualitatively on the DFT functional used, as we have found the same feature 
within GGA and also in B3LYP calculations. The presence of strong molecular resonances 
in the hybridization function makes this case different from the case of 
nanocontacts with magnetic impurities where the hybridization functions are generally 
much smoother (see Ref. \cite{jacob2009} for comparison).

\begin{figure}
  \centering
  \includegraphics[width=0.9\linewidth]{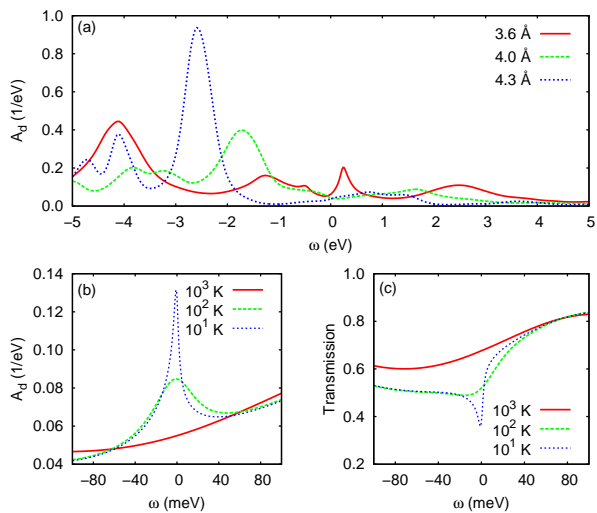}
  \caption{(Color online) 
    (a) Spectral functions of Co $3d$ shell for different $d$ at temperature
    $T\sim$1200~K. (b,c) Spectral and transmission functions of the molecule at $d=3.6$~\r{A} for different temperatures.}
  \label{oca-results}
\end{figure}

Fig. \ref{oca-results}a compares the correlated spectral functions 
$A_d(\omega)\sim{\rm Im}G_d$ of the Co $3d$ electrons for the three distances considered here at 
high temperatures on a large energy scale. The spectra vary considerably as the distance 
$d$ changes.
Most importantly, for $d$ around 3.6~\r{A} when the molecule is slightly compressed, 
a sharp temperature-dependent peak appears right at $E_F$, as can be seen 
from Fig. \ref{oca-results}b. The peak is strongly renormalized (i.e. it only carries 
a small fraction of the spectral weight) due to the strong electron-electron interactions. 

The sharp peak in the spectral function at $E_F$ that starts to develop already 
at temperatures of $kT=0.01$~eV$\approx120$~K stems from the $E_2$ channel which is the
only channel with appreciable hybridization near $E_F$, see Figs.~\ref{dft-results}b+c. 
Correspondingly, the transmission function (Fig. \ref{oca-results}c) shows a Fano-like 
feature around zero energy. A renormalized, sharp and temperature-dependent resonance in 
the spectral function at $E_F$ is commonly associated with the Kondo effect. Looking at 
the orbital occupations, we find that the $E_2$ channel that gives rise to the resonance 
for $d=3.6$~\r{A} has an occupation of about $2.8$ while the total occupation of the 
$3d$ shell is $N_d\sim7.5$. The fractional occupation numbers indicate the presence of 
valence fluctuations where the charges in the individual impurity levels fluctuate in 
contrast to the pure Kondo regime where these fluctuations are frozen.

\begin{figure}
  \begin{center}
    \includegraphics[width=0.95\linewidth]{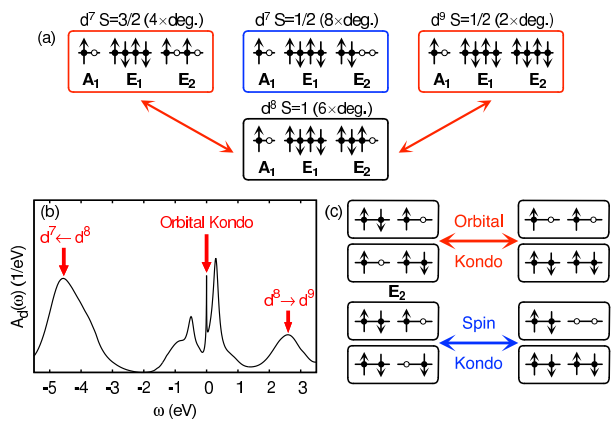}
  \end{center}
  \caption{(Color online)
    (a) Fluctuations between the atomic states that give rise to the spectral features shown in (b). 
    (b) Spectral function around $E_F$ for $d=3.6$~\r{A}. Arrows indicate spectral features arising 
    from the fluctuations between atomic states shown in (a).
    (c) Orbital (red) and spin flip (blue) fluctuations in the $E_2$ channel
    that can lead to an orbital Kondo effect and to an underscreened spin Kondo effect,
    respectively.
  }
  \label{orb-kondo}  
\end{figure}

Analyzing the atomic states of the Co $3d$-shell contributing to the ground state of the 
system we find that the principal contribution ($\sim45$\%) is an atomic state with $8$ 
electrons and a total spin of $S=1$ $(d^8, S=1)$ as shown in Fig.~\ref{orb-kondo}a. The 
total spin 1 stems from holes in the $E_2$ and $A_1$ channels. The charge fluctuations in 
the $E_2$ channel are mainly due to the contribution ($\sim17$\%) of an atomic 
$(d^7, S=3/2)$ state. There are considerably weaker contributions ($\sim4$\%) from atomic 
$(d^7, S=1/2)$ and $(d^9, S=1/2)$ states. The individual contributions of the remaining 
atomic states are very small (below 1\%) but add up to a total contribution of 34\%. 

By exclusion of individual atomic states from the calculation of the spectra we can
determine which fluctuations are responsible for the different spectral features.
We find that the fluctuations between the $(d^8,S=1)$ and the $(d^7,S=3/2)$ states are 
primarily responsible for the three spectral features close to $E_F$ including the sharp 
Kondo-like peak right at $E_F$, as illustrated in Fig.~\ref{orb-kondo}b. Also the broad 
peak around $4$~eV below $E_F$ originates from these fluctuations while the broad peak 
about $2.5$~eV above $E_F$ arises from fluctuations between the $(d^9, S=1/2)$ atomic 
state and the principal $d^8$ atomic state. The two peaks in the spectral function nearest 
to the Kondo resonance arise from the strong energy dependence of the hybridization 
function whose real part has poles just below and above $E_F$ roughly at the positions of 
these two spectral features. 

Note that the fluctuations from the $(d^8,S=1)$ to the $(d^7,S=3/2)$ states 
that give rise to the Kondo-like peak at $E_F$ actually cannot lead to a spin
Kondo effect since the spin $3/2$ of the $d^7$ state is higher than the
spin $1$ of the principal $d^8$ state. 
Instead the fluctuations between the $(d^8,S=1)$ and the $(d^7,S=3/2)$ states which give 
rise to the Kondo-like resonance at $E_F$ correspond to an {\it orbital Kondo effect} 
in the doubly-degenerate $E_2$ levels of the Co $3d$ shell as illustrated in the upper 
panel of Fig.~\ref{orb-kondo}c. Here the index labeling the two orbitals with $E_2$ 
symmetry takes over the role of a pseudo spin. In the principal $d^8$ atomic state the 
$E_2$ levels are occupied with three electrons and hence have a pseudo spin of $1/2$. 
By excitation to the $(d^7,S=3/2)$ state the electron with minority real spin and with 
some pseudo spin state is annihilated. By relaxation to the principal electronic $(d^8,S=1)$ 
state a minority real spin electron can now be created in one of the two pseudo spin 
states. Those processes that lead to a flip of the pseudo spin then give rise to the 
orbital Kondo effect and the formation of the Kondo peak at $E_F$.

The absence of a normal spin Kondo effect where the total spin 1 of the principal $d^8$ 
atomic state is screened, is easily understood on the following grounds: First, in general 
the Kondo scale decreases exponentially with increasing spin of the magnetic impurity
\cite{nevidomskyy_2009}. In addition, here the $A_1$ level does not couple at all to 
the conduction electrons around $E_F$ (no hybridization). Thus the spin $1/2$ associated
with it cannot be flipped directly through hopping processes with the conduction electron 
bath.  

On the other hand, an {\it underscreened} Kondo effect as reported in Ref. \cite{parks2010}
where only the spin $1/2$ within the $E_2$ shell is screened is also suppressed compared to 
the orbital Kondo effect due to Hund's rule coupling: Screening of the spin $1/2$ in the $E_2$ shell
can take place by fluctuations to the ($d^7,S=1/2$) state. However, the Hund's rule coupling $J$ 
favors the high spin ($d^7,S=3/2$) state over the low spin ($d^7,S=1/2$) state as can also be 
seen from the smaller weight of the latter compared to the former.
Hence the Kondo scale is considerably lower for the underscreened Kondo effect than for the orbital 
Kondo effect found here. At lower temperatures (not accessible with OCA) the two Kondo effects may in fact coexist.

We have also checked the dependence of the LDA+OCA spectra on the DCC
as well as on the interaction parameters $U$ and $J$ 
(not shown \cite{supplemental}). 
In general we find the spectra and also the Kondo peak to be 
qualitatively robust against shifts of the impurity levels in energy
over a range of several electron volts, and changes of $U$ between 3 
and 7~eV, and of $J$ between 0.6 to 1~eV. As expected for the Kondo effect
the sharp resonance stays pinned to the Fermi level when shifting the 
impurity levels in energy, and only height and width somewhat change.

Stretching the molecule by displacing the tips of the Cu nanowires 
the Kondo resonance and the concomitant Fano line shape in the 
transmission disappear for distances $d\ge4$~\r{A} (not shown).
This is accompanied by an increase of the occupation of the Co
$3d$ shell. The new regime is characterized
by a strong valence mixing between the $d^8$ and $d^9$ atomic state
of roughly equal contribution indicating that the system is now in 
the so-called mixed valence regime (see e.g. Ref.~\onlinecite{hewson}, Chap. 5).
Hence the orbital Kondo effect and the associated spectral features
can be controlled by stretching or compressing the molecule via 
the tip atoms of the Cu nanocontact.
This strong sensitivity on the molecular conformation stems from the 
sharp features in the hybridization function which change considerably  
when the molecule is stretched or compressed as can be seen from 
Fig.~\ref{dft-results}d. This peculiar behavior is qualitatively 
different from the case of the nanocontacts containing magnetic 
impurities where the hybridization functions are much smoother 
\cite{jacob2009}. 

In conclusion, we have shown for a CoBz$_2$ sandwich molecule coupled to Cu 
nanowires that the dynamic correlations originating from the strongly 
interacting Co $3d$ electrons give rise to an orbital Kondo effect in the 
doubly degenerate $E_2$ levels while a spin Kondo effect is suppressed by 
Hund's rule coupling.
Due to the sensitivity of the electronic correlations on the molecular 
conformation it is possible to control the appearance of the orbital 
Kondo effect by stretching and compressing the molecule with the tips 
of the Cu leads.
It should be possible to prepare a setup similar to the one 
considered here in an actual experiment and measure the orbital 
Kondo effect, e.g. by contacting a CoBz$_2$ molecule deposited on 
a Cu(111) surface with a scanning tunneling microscope.
Finally, we mention that we have also explored the case of a NiBz$_2$ sandwich.
In this case we do not find a Kondo effect. Instead the system is in the mixed valence
regime characterized by a strong peak with weak temperature dependence close to but
not at $E_F$ \cite{supplemental}.

We thank K. Haule for providing us with the OCA impurity solver and for helpful 
discussions. 
Support from SFB 668, LEXI Hamburg and ETSF are acknowledged. 
MK gratefully acknowledges the hospitality of the Max-Planck-Institute 
in Halle (Saale).

\bibliography{thebib}

\end{document}